\documentclass[%
 reprint,
 superscriptaddress,
 amsmath,amssymb,
 aps,
 pre
floatfix,
]{revtex4-1}

\usepackage{graphicx}
\usepackage{dcolumn}
\usepackage{bm}
\usepackage{subfigure}

\begin{document}

\title{Conservation of population size is required for self-organized criticality in evolution models}

\author{Yohsuke Murase}
\affiliation{RIKEN Center for Computational Science, 7-1-26, Minatojima-minami-machi, Chuo-ku, Kobe, Hyogo, 650-0047, Japan}

\author{Per Arne Rikvold}
\affiliation{Department of Physics, Florida State University, Tallahassee, FL 32306-4350, USA}
\affiliation{PoreLab, Department of Physics, University of Oslo, P.\ O.\ Box 1048, Blindern, 0316 Oslo, Norway}
\altaffiliation[]{Current address.}


\date{\today}

\begin{abstract}
We study models of biological evolution and investigate a key factor to yield self-organized criticality (SOC).
The Bak-Sneppen (BS) model is the most basic model that shows an SOC state, which is developed based on minimal and plausible assumptions of Darwinian competition.
Another class of models, which have population dynamics and simple rules for species migrations, has also been studied.
It turns out that they do not show an SOC state although the assumptions made in these models are similar to those in the BS model.
To clarify the origin of these differences and to identify a key ingredient of SOC, we study models that bridge the BS model and the Dynamical Graph model, which is a representative of the population dynamics models.
From a comparative study of the models, we find that SOC is found when the fluctuations of the number of species $N$ are suppressed, while it shows off-critical states when $N$ changes according to its evolutionary dynamics.
This indicates that the assumption of the fixed system size in the BS model plays a pivotal role to drive the system into an SOC state, and casts doubt on its applicability to actual evolutionary dynamics.
\end{abstract}

\pacs{Valid PACS appear here}
\maketitle


\section{Introduction}\label{sec:intro}

The Bak-Snpeppen (BS) model \cite{bak93:_punct_equil_and_critic_in} is one of the simplest models of biological evolution and has attracted much attention as it shows self-organized criticality \cite{PhysRevE.53.414,bak-pnas1995,sneppen-pnas1995,flyvbjerg93:_mean_field_theor_for_simpl,moreno2002bak,fraiman2018bak,de1998high,dickman2000paths}.
The self-organized criticality (SOC) denotes that the system goes into a critical state without the tuning of a parameter, showing large extinction avalanches \cite{newman2003me} and intermittent dynamics \cite{eldredge1972pea,gould1977pet}, whose statistics are characterized by power laws \cite{bak-pnas1995,Sole:1997hi}.
It has been suggested that the mass extinctions and punctuated equilibria found in the history of life are consequences of the SOC of the ecosystem, though counterarguments also exist \cite{Kirchner:1998wa,Sole-1996,alroy2008dynamics}.
The BS model is developed based on the minimal and plausible assumptions of Darwinian competition:
A system of a fixed number of species $N$ is updated by successive extinctions of the least fit species and migrations of new species whose fitness and interactions are drawn randomly.
By repeating this simple process, the system self-organizes into a critical state, seemingly implying that successive exclusion of unfit species always drives the system into the critical state.

On the other hand, another class of models for biological evolution has also been studied.
This class of models has population dynamics of species (either at species level or at individual level) and rules for introducing new species, aiming at bridging ecological and evolutionary time scales.
This class of models includes the Tangled-Nature models \cite{CHRISTENSEN:2002yq,PhysRevE.66.011904,0305-4470-36-4-302,rikvold2003pea,Rikvold:2007lr,cairoli2014forecasting}, the web-world model \cite{Drossel:2004fj,drossel01:_influen_of_predat_prey_popul}, the scale-invariant model \cite{shimada2007simple,shimada-arob2002}, and the replicator model \cite{tokita-tpb2003}.
Each of these models has its own functional form of the population dynamics and there seems to be no de-facto standard.
For instance, some of them assume predator-prey interactions between species \cite{Rikvold:2007lr,shimada-arob2002}, while others allow more general inter-species interactions \cite{PhysRevE.66.011904,rikvold2003pea}.
Interestingly, the statistical properties (such as the extinction-size distribution, the species lifetime distribution, and the intermittency of the time series) obtained for these models seems to be categorized into a few classes despite their wide variety in network size and network topology \cite{murase2010random,rikvold-2005}.
These statistical properties are dependent on a few key factors of the models, such as the introduction of a genome space \cite{murase2010random} or demographic stochasticity \cite{murase2010effects}.
For instance, the Tangled-Nature models assume that new species appear in the system by mutations of existing species, each of which is represented by a single bit flip in an $L$-bit ``genome'' string.
These models show intermittent evolutionary dynamics, characterized by quasi-stable states interrupted by sudden and active reorganization of species.
Among these various models, the simplest rule of introducing new species is the migration rule \cite{murase2010random}.
In the migration rule, new species, whose links as well as their weights are randomly drawn from a given probability distribution, are added to the system at a constant rate.
Within this setting, a simple exponential extinction size distribution and ``skewed'' species-lifetime distribution are robustly found irrespective of the functional form of the population dynamics, meaning no sign of SOC is found.
Although this class of models seems to be based on plausible assumptions similar to the BS model, the latter shows SOC behavior while the former does not.
The aim of this paper is to understand the origin of these differences and to identify the key element in the model definitions that causes the SOC behavior in order to deepen the understanding of evolutionary dynamics and SOC phenomena in general.

In this paper, we focus on a comparison between the BS model and the Dynamical Graph (DG) model \cite{murase2010simple,shimada2014universal,murase2015universal}, which is a simplified variant of the migration models.
The DG model is a simplistic model developed based on a minimal set of plausible assumptions in order to analyze the underlying mechanisms for the migration population dynamics models.
In the DG model, an ecosystem is represented by a weighted directed network.
The species' fitness is defined as the sum of the incoming link weights, and a species survives until its fitness becomes negative.
(The detailed model definition is given in the next section.)
Even though the model no longer has population dynamics equations, it reproduces the statistical properties of the corresponding population dynamics models, i.e., the model does not show any SOC behavior.
Thanks to its simplicity, the mechanisms for the skewed species-lifetime distribution becomes clear and its functional form turns out to be a stretched exponential function with exponent $1/2$ \cite{murase2010simple,murase2015universal}.
We use this model as a representative of the non-SOC models since it is as simple as the BS model.

We propose that the fixed number of species is a key building block for the SOC.
In the DG model, the number of species may change with time through the evolutionary process, while the BS model has a fixed number of species.
In the sandpile SOC model \cite{bak1987self}, it is known that conservation of the number of particles is a necessary factor for the SOC \cite{manna1990cascades,ghaffari1997nonconservative,tsuchiya2000proof}.
Although an analogous quantity for the evolutionary model is not known, we conjecture that conservation of the number of species is a necessary condition for SOC.
To test this conjecture, we propose a model which suppresses the fluctuations of the number of species by a control parameter.
The model is equivalent to the DG model in one limit of the parameter, while the model has a fixed number of species like the BS model in the other limit.
We will see how SOC behaviors emerge as this parameter changes to suppress the population fluctuations.

This paper is organized as follows.
In the next section, we review the DG model and the BS model in detail, discussing the key differences between these two models.
In Section 3, we propose a generalized model which incorporates the DG model and the BS model as special cases, and we show the simulation results for this model.
The last section is devoted to summary and discussions.

\section{A review of the models}\label{sec:model}

\subsection{Dynamical Graph model}

In the DG model, a directed weighted network is considered, where nodes and links represent species and interspecies interactions, respectively.
The fitness of each species (node) is defined as the  sum of its incoming links:
\begin{equation}\label{eq:fitness}
  f_i = \sum_j a_{ji},
\end{equation}
where $a_{ji}$ is the weight of the link from node $j$ to node $i$.
The species goes extinct if its fitness becomes less than the threshold $f_{\rm th} = 0$ and survives otherwise.
At each time step, the network is updated by migration of a new species and the subsequent extinctions.
When a new species migrates, links between the new species and extant species are made with probability $c$ for each direction, and its weight is randomly drawn from a Gaussian distribution with mean $0$ and standard deviation $1$.
(See Fig.~2 in~\cite{murase2010simple} or Fig.~1 in~\cite{shimada2014universal}.)
Because of the introduction of the new species, some of the existing species, including the new one, may have a negative fitness.
The species with the lowest negative fitness in the system is removed.
Because of the extinction of this species, the fitnesses of the neighboring species change accordingly.
The fitnesses are calculated again, and the species having the lowest fitness is removed if its fitness is negative.
The removal of species continues until all the species in the system have a positive or zero fitness, causing an avalanche of extinctions.
Since extinctions may or may not happen, the number of species $N$ changes with time.
In this setting, the system reaches a statistically stationary state after a sufficiently long initialization period starting from an empty network.

We show the statistical properties of the DG model in Fig.~\ref{fig:bsdg}(a).
The distribution of the fitness $P(f)$ is shown in Fig.~\ref{fig:bsdg}(a-1). It is positive only for $f\geq 0$ by model definition.
The extinction size for this model is defined as the number of the species that went extinct in one time step.
The distribution of extinction size follows a simple exponential function as shown in Fig.~\ref{fig:bsdg}(a-2).
The interval between extinctions, which we define as the number of steps between consecutive extinctions (or the number of consecutive steps having no extinction), shows an exponential distribution as well (Fig.~\ref{fig:bsdg}(a-3)).
The time series of the number of species for this model shows a $1/f^2$ power spectral density, which indicates the dynamics is characterized by a simple Ornstein-Uhlenbeck process.
All these statistics indicate that the extinction events occur randomly, characterized by a Poisson process.
A non-trivial aspect of this model is found in the lifetime distribution of species.
The lifetime of a species is defined as the duration between its immigration and extinction.
The species lifetime distribution $P(L)$ follows a stretched exponential function, $P(L) \propto \exp{\left(-(L/L_0)^{\alpha}\right)}$, with the exponent $\alpha$ close to $1/2$.
This functional form is explained by the modified Red-Queen hypothesis, where an age-independent and $N$-dependent mortality is assumed~\cite{murase2010simple,murase2015universal}.

\subsection{Bak-Sneppen model}

Contrary to the DG model, the BS model assumes that an ecosystem has a fixed number of species $N$ and each species $i$ has a single fitness value $f_i$.
For each step, the species having the minimum fitness is removed from the system (extinction) and immediately replaced with a new one (migration).
The fitness of the new species is uniformly randomly drawn from $[0,1]$, and the fitness of its neighbors are also updated to a randomly assigned value, modeling the interactions with the new species.
Following Ref.~\cite{bak93:_punct_equil_and_critic_in}, we assume that the fitness barrier that must be overcome before a species can go extinct is proportional to its fitness.
This leads to the time before extinction of the least fit species, $f_{\rm min}$, and its replacement with a new species being of the form $\tau_{\rm ext}(f_{\rm min}) = t_0\exp ( f_{\rm min}/f_0 )$, where $t_0$ and $f_0$ are constants.
Here, $f_0$ must be sufficiently small to produce a broad region of critical scaling in the temporal dynamics.
In the following, we use $t_0=1$ without loss of generality.
So the actual time scale proceeds by $\tau_{\rm ext}(f_{\rm min})$ in each step.
Although species were assumed to be aligned along a one-dimensional lattice in the first paper where the BS model is originally proposed~\cite{bak93:_punct_equil_and_critic_in}, we mainly focus on the random-neighbor (or mean-field) version of the model\cite{flyvbjerg93:_mean_field_theor_for_simpl}, in which a new species interacts with randomly chosen species.
This simplification is instrumental not only for analytical treatment~\cite{flyvbjerg93:_mean_field_theor_for_simpl,fraiman2018bak} but for comparison with the DG model.

\begin{figure*}
  \begin{center}
  \includegraphics[width=.99\textwidth]{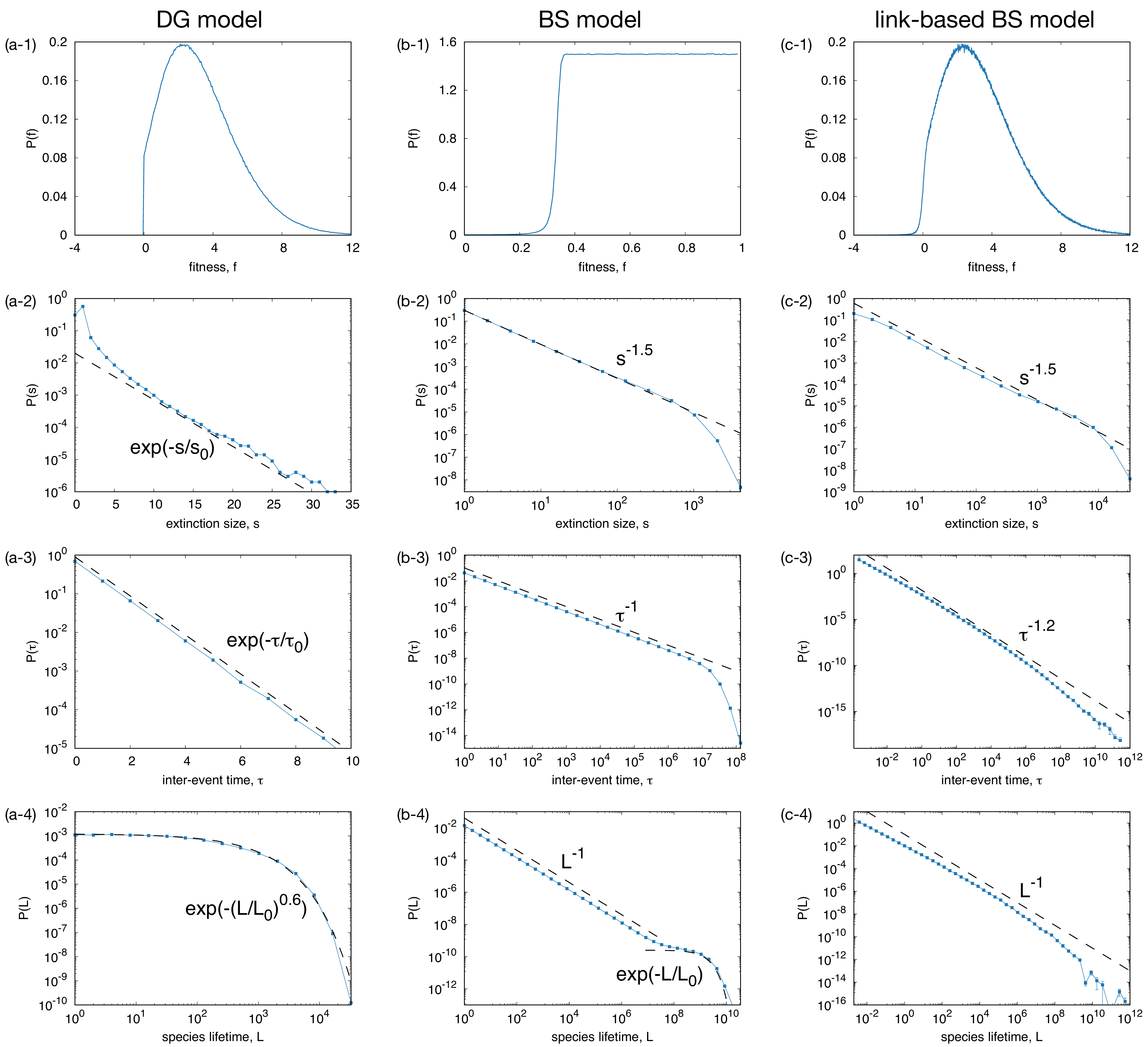}
  \caption{ \label{fig:bsdg}
  A summary of the statistical properties for (left) the DG model, (center) the random-neighbor BS model, and (right) the link-based BS model.
  From top to bottom, the probability distribution of fitness $P(f)$, the extinction size distribution $P(s)$, the inter-event time distribution $P(\tau)$, and the species-lifetime distribution $P(L)$.
  \\
    (Left) The parameters for these plots are $c=0.01$. The statistics are taken for $10^7$ steps with a $5\times 10^5$ initialization period.
  The histogram in (a-1) is taken with a bin size of $0.01$.
  Note that the horizontal axes in (a-2) and (a-3) are on a linear scale.
  Thus these are well fitted by exponential functions.
  Fitting curves are shown in Figs.~(a-2), (a-3) and (a-4) by dashed curves as guides to the eyes, whose functional forms are shown in each figure.
  \\
  (Center) The parameter $f_0=0.02$ and the number of interactions $m=2$ are used. (b-1) The histogram is taken with a bin size of $0.01$. It is near zero for $0 < f < f_{\rm th}$ while it shows a flat profile for $f_{\rm th} < f < 1$, where an analytically obtained threshold $f_{\rm th}=1/3$ \cite{fraiman2018bak,flyvbjerg93:_mean_field_theor_for_simpl}. The gray dashed curves shown in (b-2), (b-3), (b-4) are power laws and an exponential curve shown as guides to the eyes. The functional form of the fitting curves are shown in each figure.
  The simulations are conducted for $10^7$ steps with a $65536$ steps initialization period. The results are averaged over $5$ independent runs and their statistical errors are smaller than the symbol size.
  \\
  (Right)
    The statistical properties for the link-based BS model for the parameters $N=1000$, $c=0.01$, and $f_0=0.02$., and $f_{\rm th}=0.2$.
    For all these figures, the results are averaged over $5$ independent runs, each of which runs for $10^6$ time steps with $10^5$ initialization steps.
    In (c-1), the histogram is taken with a bin size of $0.01$.
    In Fig.~(c-2), the extinction size distribution calculated with the threshold $f_{\rm th}=0.2$ is shown.
    In Figs.~(c-2), (c-3), and (c-4), the statistics shows power laws although there are slight deviations.
    The fitting curves are shown as gray dashed lines, whose functional forms are shown in each figures.
  }
  \end{center}
\end{figure*}

A statistically stationary state is obtained after a sufficient number of iterations.
For comparison, we show the results for the BS model in Fig.~\ref{fig:bsdg}(b).
The distribution of $f$, $P(f)$, is shown in Fig.~\ref{fig:bsdg}(b-1).
It shows a profile similar to a step function: a positive plateau is found for $f_{\rm th} < f < 1$, while it drops to an infinitesimal value for $0 < f < f_{\rm th}$.
When we see the temporal dynamics of the extinction events, the dynamics shows an intermittent behavior characterized by a power-law inter-event time distribution (Fig.~\ref{fig:bsdg}(b-3)), reminiscent of punctuated equlibria.
Here, the inter-event time is defined as the duration between two consecutive extinctions, i.e., $\tau_{\rm ext}$.
The extinction size is defined as the number of consecutive extinctions whose $f_{\rm min}$ is smaller than $f_{\rm th}$.
From the viewpoint of the time series, the extinction size is defined as the size of the ``correlated bursts'' \cite{karsai2012universal,karsai2018bursty}.
We obtain the clusters of events, or the bursty train, by splitting the event sequence by interval $\Delta = \tau_{\rm ext}(f_{\rm th})$, and the number of events in each group is regarded as the extinction size.
In Fig.~\ref{fig:bsdg}(b-2), the extinction size is shown. It is well fitted by a power law of exponent $-3/2$.

The species lifetime can be defined as the time between a species' appearance and its extinction although it has not been studied in the literature to our knowledge.
From our simulations, as shown in Figs.~\ref{fig:bsdg}(b-4), the species lifetime distributions shows a power law followed by an exponential distribution.
The exponent for the power law part is $-1$, which is same as that for the inter-event time distribution.
The typical time scale for the latter part is approximately $\tau_{\rm ext}(f_{\rm th})$.
Since the neighbor species and its renewed fitness is randomly selected, each species always has a positive finite probability of having a fitness below the threshold, which leads to its extinction.
Thus, the species lifetime distribution has an exponential part although the overall profile is $L^{-1}$.

\subsection{Key differences}\label{subsec:key_differences}

We summarize the key differences in the model definitions between the BS and the DG models as follows.
\begin{enumerate}
\item
  While the number of species in the DG model fluctuates around its equilibrium value, the BS model has a fixed number of species which is given as a model parameter.
\item
  The extinction threshold $f_{\rm th}$ is predefined in the DG model. On the other hand, in the BS model, the threshold of the fitness is self-organized as a result of immigrations and extinctions.
\item
  The time required for immigrations and extinctions are different.
  In the DG model, immigrations occur regularly at each time step. The extinction immediately happens when a species has a negative fitness.
  By contrast, in the BS model, the time required for an extinction is dependent on the fitness value, which is followed by an instantaneous immigration of a new species.
\item
  The fitness is defined as the sum of the incoming interspecies interactions in the DG model while the fitness for the BS model is defined as an attribute of a species.
\end{enumerate}
The first three factors are not mutually exclusive but dependent on each other.
These factors are reminiscent of the difference between canonical and grand canonical ensembles.
The canonical ensemble, where the number of particles is fixed and the chemical potential is obtained as an outcome, may be regarded as similar to the BS model, where the number of species is fixed and the threshold is obtained as an outcome.
Similarly, the grand canonical ensembles, where the number of particles is obtained given the chemical potential, is analogous to the DG model, where the number of species is obtained given the extinction threshold.
Thus, it is not easy to incorporate only one of these factors.

The fourth factor, however, is independently testable by modifying the BS model so that the fitness of the species is defined as the sum of the interactions.
In the following section, we will first show that the results for this ``link-based'' BS model remain qualitatively the same as those for the original BS model, implying that Factor $4$ is not the key ingredient of the SOC.
We will then generalize the model to interpolate between the link-based BS model and the DG model to see the effects of Factors $1$, $2$ and $3$.

\section{Results}

\subsection{link-based BS model}

We first investigate a variant of the BS model, where the fitness of a species is defined as the sum of the inter-species interactions.
More specifically, the ecosystem is represented by a weighted directed network whose nodes and links represent species and interactions between them, respectively.
At each time step, the species with the minimum fitness as well as its links are removed.
Similar to the BS model, the extinction is followed by an immediate immigration of a new species, whose incoming and outgoing links are made with probability $c$ and their weights are randomly independently drawn from a Gaussian distribution with mean $0$ and variance $1$.
Thus, the fitness of the other species undergoes changes both by the eliminations of the links and by the introduction of the new links.
The time required for the extinction to occur is the same as that for the BS model, i.e., $\tau_{\rm ext}(f) = \exp ( f/f_0 )$.

This modification does not alter the results significantly from the BS model.
See Figs.~\ref{fig:bsdg}(c) for the results.
As shown in Fig.~\ref{fig:bsdg}(c-1), the fitness distribution shows a sudden drop at a certain threshold $f_{\rm th} \approx 0$ similarly to the BS model although the shape of the distribution and the threshold value are different.
The extinction size is defined similarly to the BS model.
The distribution of the extinction size $P(s)$ shows a power law with exponent $-3/2$ as shown in Fig.~\ref{fig:bsdg}(c-2), which is the same as that for the BS model.
In Fig.~\ref{fig:bsdg}(c-3), the inter-event time distribution $P(\tau)$ is shown.
It shows an approximate power law with exponent around $-1.2$, although the profile is slightly concave on the log-log scale.
The lifetime distribution, shown in Fig.~\ref{fig:bsdg}(c-4), is also approximated by a power law with exponent $-1$.
Thus, we conclude that the model shows SOC behavior, which is essentially the same as the BS model. 
The link-based definition of species fitness does not alter the picture fundamentally.

\subsection{generalized model}

We propose the following model in order to study the effect of the key differences 1, 2, and 3 discussed in Section~\ref{subsec:key_differences}.
We generalize the DG model, incorporating some ingredients of the BS model.
The system is represented as a weighted directed network and a species' fitness is defined as the sum of its incoming links.
Following the assumption made in the BS model, we assume that the time required for an extinction of a species depends on the fitness as $\tau_{\rm ext}(f) = \exp (f / f_0)$, where $f_0$ is a constant determining typical scale.
In addition to this, we also consider the time between migration events in order to control the number of species.
We assume that the time required for a migration depends on the current number of species and a parameter $\mu$ which controls the fluctuations in $N$: $\tau_{\rm mig}(N) = \exp \left( \mu (N-N_{0}) / f_0 \right)$.
Here, $N_0$ is an input parameter of the model, denoting the expected number of species.
When $\mu = 0$, the migration time is always $\tau_0=1$, irrespective of the current number of species.
There is no external force controlling the number of species, which corresponds to the situation for the DG model.
When $\mu$ is large enough, on the other hand, the number of species is controlled by an external force.
Migrations occur frequently when $N < N_{0}$, putting species into the system more often, while migration occurs less frequently when $N > N_0$.
Thus, the number of species is under stronger control yielding smaller fluctuations in the number of species around $N_0$.
In the limit of $\mu \to \infty$, immigration occurs immediately when $N<N_0$, while it never occurs when $N>N_0$.
This results in a situation that immigrations and extinctions occur one after the other as in the BS model where $N$ is fixed.

The other rules are kept the same as the DG model.
When a migration of species occurs, a species whose interactions are randomly determined is added to the system.
For each possible link between existing species, we make a link with probability $c$ and the weight of the links are drawn from a Gaussian distribution with mean $0$ and variance $1$.
The algorithm to update the system is summarized as follows.
\begin{enumerate}
  \item Find the species with the minimum fitness in the system, $f_{\rm min}$.
  \item Calculate $\tau_{\rm ext}(f_{\rm min})$ and $\tau_{\rm mig}(N)$.
    \begin{itemize}
      \item If $\tau_{\rm ext} < \tau_{\rm mig}$, an extinction of the least fit species occurs, i.e., the species is removed from the system.\\
      The current time $t$ is increased by $\tau_{\rm ext}$.
      \item If $\tau_{\rm ext} \geq \tau_{\rm mig}$, a migration of a new species occurs.\\
      The current time $t$ is increased by $\tau_{\rm mig}$.
    \end{itemize}
\end{enumerate}

\begin{figure}
\begin{center}
\subfigure{
  \includegraphics[width=.48\textwidth]{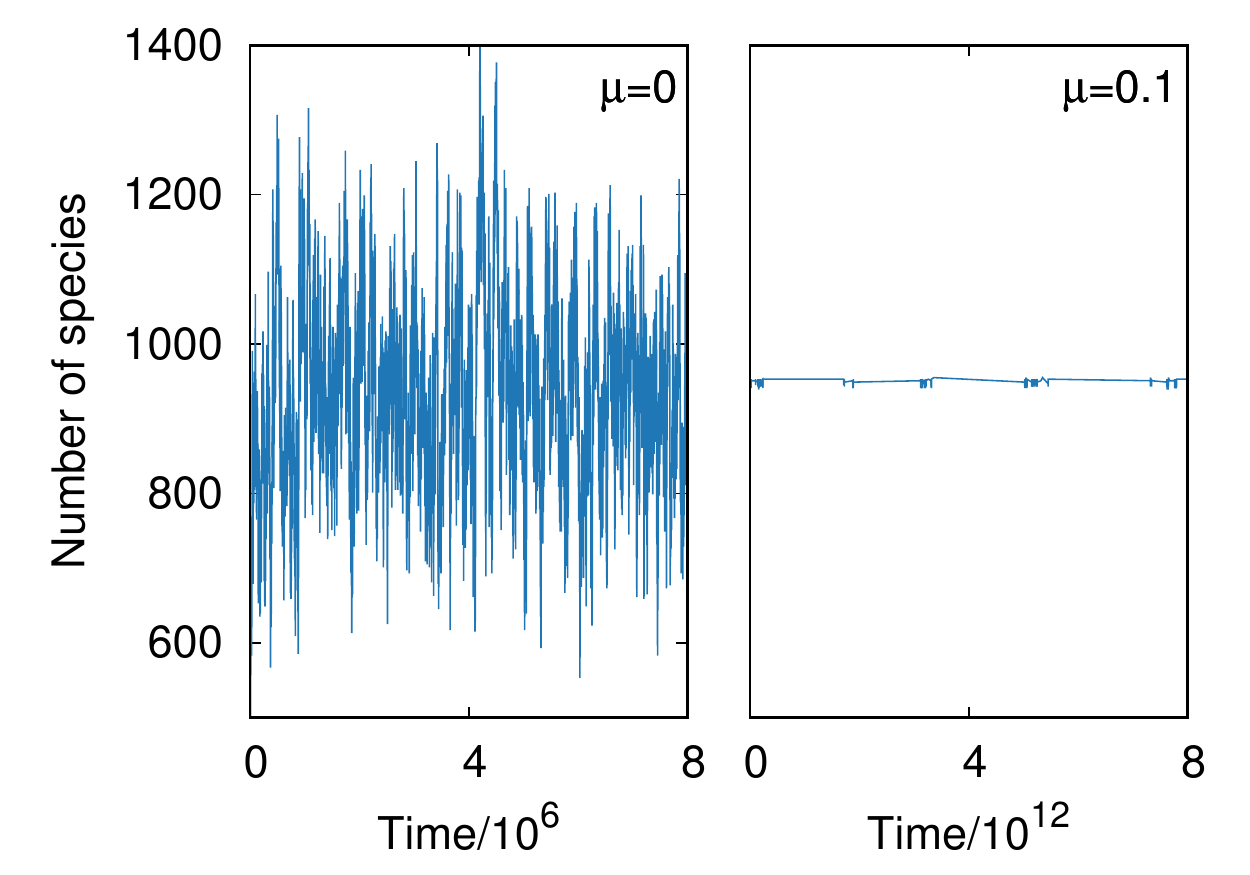}
  }
 \subfigure{
  \includegraphics[width=.48\textwidth]{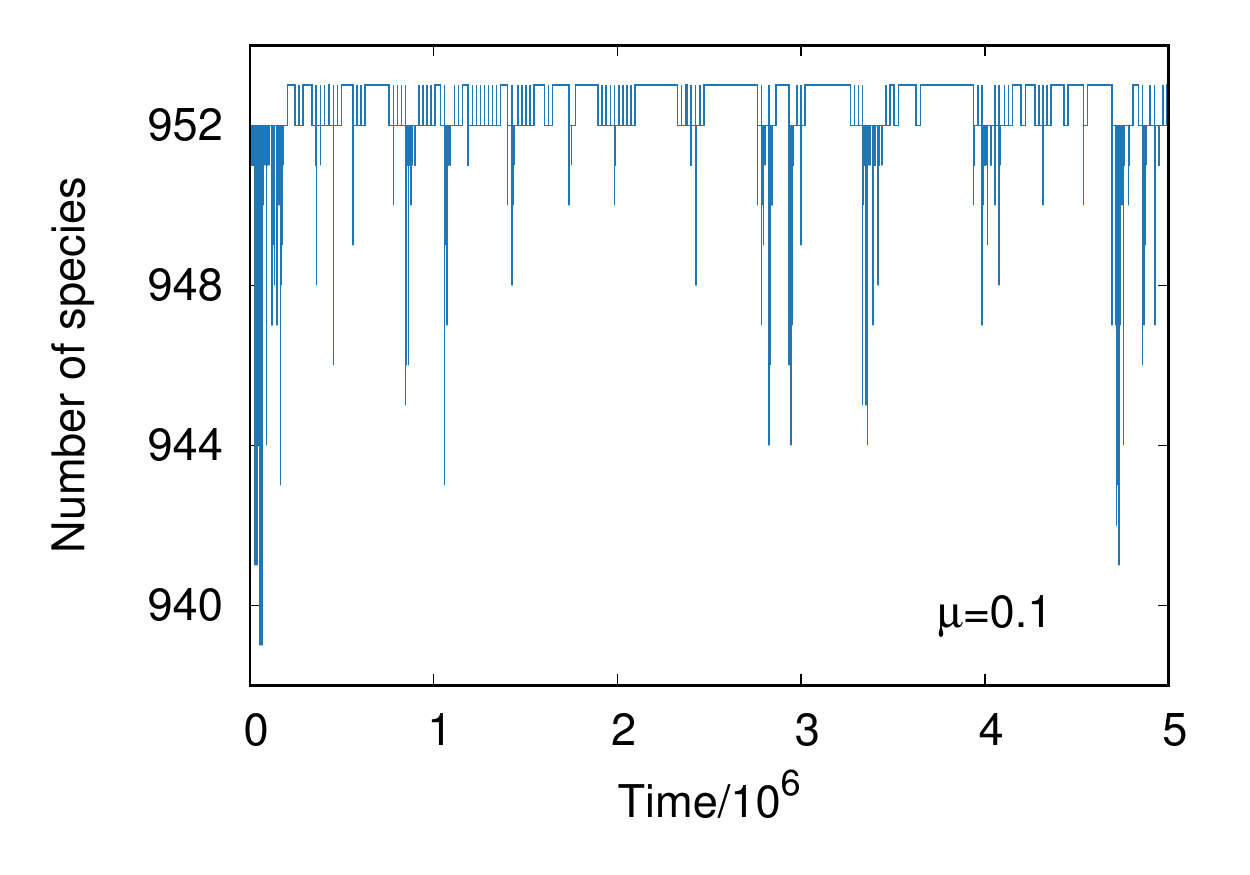}
 }
\end{center}
\caption{ \label{fig:gen_timeseries}
  (Top) Time series of the number of species for the generalized model with $\mu=0$ and $\mu=0.1$.
  The connection probability $c$ is $0.01$.
  The $x$-axis denotes the actual time scale, not the number of extinction or migration events.
  These results are for single Monte Carlo runs and are sampled for every $1024$ steps.
  The parameters $f_0=0.02$ and $N_0=950$ are used.
  When $\mu = 0$, the number of species fluctuates around its equilibrium value.
  As $\mu$ is increased, the fluctuation are suppressed and converge around the targeted value $N_0$ as shown in the right figure.
  (Bottom) A magnified plot of the time series for $\mu=0.1$.
  The time series shows intermittency. Quasi-stationary states are interrupted by active periods, where a large number of extinctions and migrations happen in a short period of time.
}
\end{figure}

The model has a correspondence to the DG model when $\mu=0$.
This is because species with negative fitness are removed during one migration as $\tau_{ext} < \tau_{mig}$ for $f_{min} < 0$. As long as there is a species with negative fitness, extinctions continue.
When all the species have a positive or zero fitness value, the system accepts an immigrant.
Although the actual time scale may be different from the DG model, the long-time behavior is approximately the same for a sufficiently small $f_0$ since $\tau_{\rm ext}$ is negligibly small.
When $\mu \to \infty$, the model has a correspondence to the BS model.
If $N > N_0$, extinction of the least fit species always occurs since $\tau_{mig} \to \infty$.
If $N < N_0$, on the other hand, migration of a new species instantly occurs.
Thus, migration alternates with extinction, keeping the number of species nearly constant around $N_0$.
In other words, the least fit species are removed and immediately replaced by a new species.
This dynamics is essentially similar to the BS model.

Typical time series for this model with $\mu=0$ and $0.1$ are shown in Fig.~\ref{fig:gen_timeseries}.
As expected, the number of species is controlled when $\mu = 0.1$ while it fluctuates widely for $\mu=0$.
Hereafter, the parameter $N_0$ is set to $950$, which is comparable to the average number of species for $\mu=0$, in order to keep the number of species when changing $\mu$ and eliminate any side-effect caused by a change in $N$.

\begin{figure*}
\begin{center}
  \includegraphics[width=.92\textwidth]{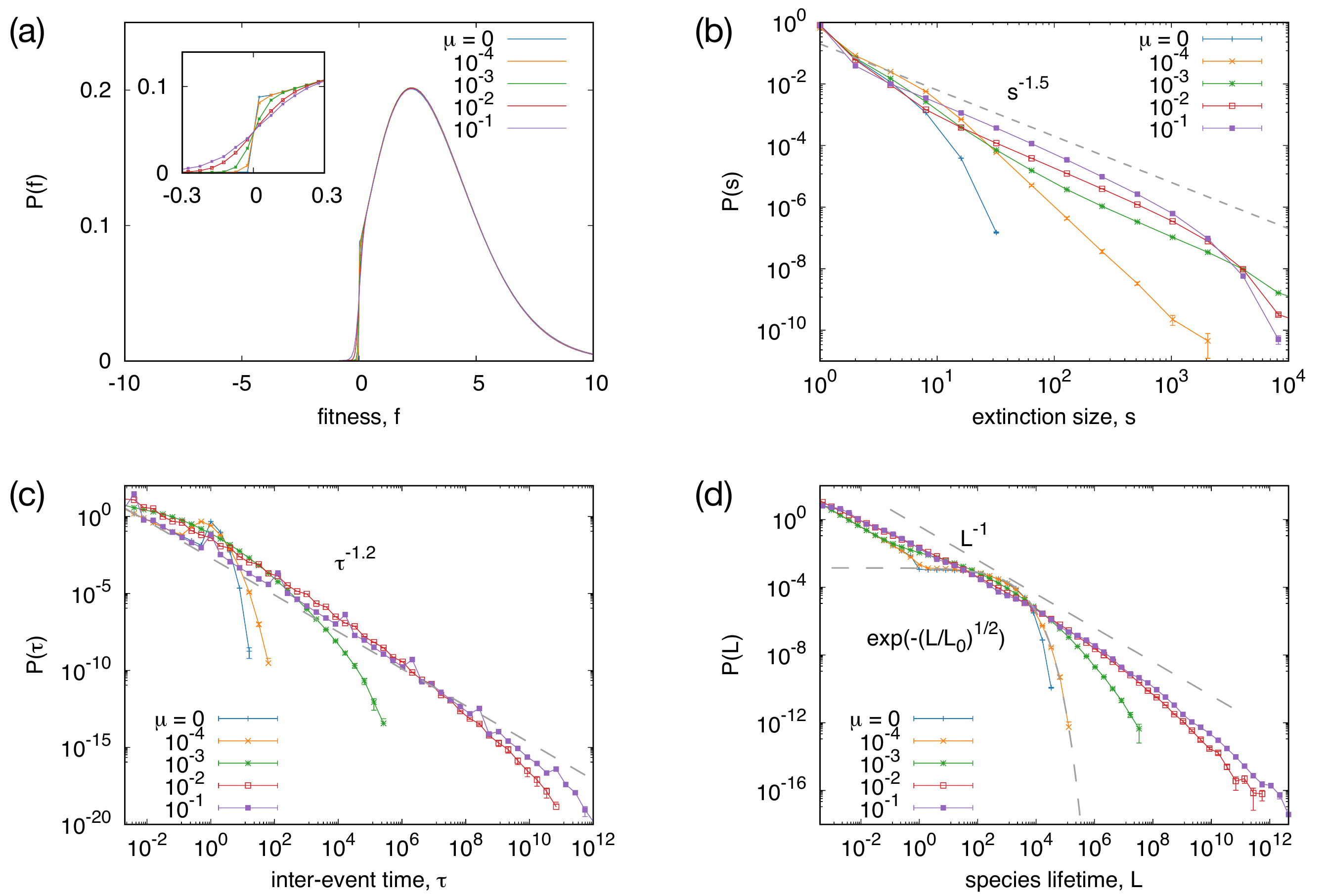}
\end{center}
\caption{ \label{fig:gen_model}
Simulation results for the generalized model with various values of $\mu$.
(a) The probability distribution of fitness. Similar to the link-based BS and DG models, a cutoff near $f=0$ is observed irrespective of $\mu$.
In the inset, the same data are shown with a magnified view near $f=0$.
(b) The distribution of the extinction size for various $\mu$. The threshold $f_{\rm th}=0$ is used.
When $\mu$ is zero, the distribution decays exponentially as in the DG model.
As $\mu$ is increased, a transition to a power law is observed. The gray dashed line is a plot for a power law with exponent $-1.5$ as in the BS model.
(c) The inter-event time distribution. The inter-event time is defined as the period between two consecutive extinctions.
(d) The species lifetime distribution. When $\mu$ is small, an initial power law decay followed by a stretched exponential curve is observed.
When $\mu$ is large enough, an approximate power law is observed.
The dashed gray curves are a stretched exponential function and a power law.
The parameters $c=0.01$, $N_0=950$ and $f_0=0.02$ are used.
Each simulation runs for $4 \times 10^7$ steps in addition to $6\times 10^5$ initialization period which are not included in the statistics.
The data are averaged over $5$ independent runs. Error bars denote the standard errors.
}
\end{figure*}

In Fig.~\ref{fig:gen_model}(a), the probability distribution of $f$ for various values of $\mu$ are shown.
As shown in the figure, the distribution has positive values for $f>0$ while it drops quickly for $f<0$, similar to the link-based model.
This profile does not change significantly by changing $\mu$ and the threshold value seems to be around zero.

We define the extinction size as the number of consecutive extinctions whose interval is less than $\Delta = \exp\left(f_{\rm th}/f_0\right)$.
This definition is consistent with those in the DG and the BS models when $\mu=0$ and $\mu \to \infty$, respectively.
Using $f_{\rm th}=0$, the extinction size is calculated for various values of $\mu$ and is shown in Fig.~\ref{fig:gen_model}(b).
When $\mu=0$, the extinction size distribution decays exponentially.
As we increase $\mu$, the tail becomes heavier and an overall power-law distribution is found for a sufficiently large $\mu$.
When $\mu=0.1$, the distribution is approximately fitted by a power law with exponent $-1.5$.
Although there is a slight deviation from a power law for large $s$ ($\sim 10^4$) and a non-monotonic dependence on $\mu$ is found, it may be due to a finite-size effect since the extinction size is the same order as $N_0$ when it deviates from the power law.

The distribution of inter-event time, which is defined as the period between two consecutive extinctions, is shown in Fig.~\ref{fig:gen_model}(c).
When $\mu=0$, the distribution consists of an initial power-law decay for $\tau < 1$ and an exponential decay for $\tau \geq 1$.
The former part corresponds to the extinctions within a single migration event, which is represented as the data point for $\tau=0$ in Fig.~\ref{fig:bsdg}(b-3).
For $\tau \geq 1$, the curve decays exponentially. It is consistent with the results for the DG model.
The tail of the distributions gets much broader and closer to a power law as we increase $\mu$.
In the figure, a line indicating a power law $\tau^{-1.2}$ is shown as a gray dashed line.
The distribution is fitted fairly well by a power law although the curve is slightly concave in a log-log plot, indicating the time series are bursty.
(One may also find the burstiness from Fig.~\ref{fig:gen_timeseries}.)
\footnote{When $\mu=10^{-1}$, there are small spikes for every two decades. These come from the discreteness of the $\tau_{\rm mig}$.}

Similar transient behavior is also observed in the species lifetime distribution $P(L)$.
As shown in Fig.~\ref{fig:gen_model}(d), the profile changes significantly with $\mu$.
When $\mu$ is small enough, the curve shows a quick decay while it has heavy tails for larger $\mu$.
The curves for small $\mu$ are well fitted by stretched exponential functions with exponent close to $1/2$, which is similar to the DG model.
As $\mu$ increases, the curve becomes closer to a power law.
For $\mu = 10^{-1}$, the distribution is approximated by a power law $L^{-1}$.

Therefore, all these statistics indicate a crossover from a non-SOC behavior to SOC with increasing $\mu$, showing that the constraint on the number of species is an essential factor for the emergence of SOC.

\section{Summary and Discussion}\label{sect:summary}

In summary, we formulated and studied a model that bridges the DG model and the BS model in order to identify a key factor for generating the SOC phenomena in the biological evolution model.
By a comparative study of a model that controls the fluctuations of $N$, we found that the constraint on the number of species significantly alters the model behavior.
When $N$ is fixed, an extinction of a species is followed by immediate replacement by a new species, i.e., the system is under a high pressure of potential new species trying to migrate into it.
Such immediate introductions of new species is a necessary condition to keep the system in a critical state.
If this condition is not met, the system goes to an off-critical state as $N$ decreases, preventing critical avalanches of extinctions.

Although the BS model has been used to explain the origin of punctuated equilibria or large-scale extinctions, its applicability is questionable as the conservation of the system size is not satisfied in general.
To realize an SOC state, it is required to have some external mechanism to maintain the system size at a constant level in addition to intrinsic Darwinian evolutionary competition, indicating that the BS model is valid only in limited situations.
Future studies are needed to explore other possibilities for modeling evolutionary dynamics.

In this paper, we limited ourselves to the simplest cases, in which new species have completely random phenotype.
Clearly, it is an idealized assumption and a different rule for introducing new species may alter the story as well.
Actually, it is known that qualitatively different results are obtained for the ``mutation'' Tangled-Nature models \cite{murase2010random}.
In these models, a species represented by an $L$-bit genome sequence may mutate to another species by flipping one of the bits.
The models yield a bursty time series characterized by $1/f$-fluctuations and power-law species-lifetime distributions.
It would be interesting to simplify these mutation models as in the DG model in order to understand the origin of the burstiness.
Other models have also been studied that show SOC states while allowing the system size $N$ to change\cite{watanabe2015fractal,slanina1999extremal,guill2008emergence}.
Each of these has its own rules for introduction and eliminations of entities, and a unified understanding is still lacking.
Further studies in this direction are expected, and we believe the present study provides a foundation for better understanding of more complex models.

\begin{acknowledgments}
YM acknowledges support from MEXT as ``Exploratory Challenges on Post-K computer (Studies of multi-level spatiotemporal simulation of socioeconomic phenomena)'' and from Japan Society for the Promotion of Science (JSPS) (JSPS KAKENHI; grant no. 18H03621).
PAR gratefully acknowledges hospitality at the RIKEN Center for Computational Science, and in the Department of Physics, Graduate School of Science, University of Tokyo. 
\end{acknowledgments}

\bibliographystyle{apsrev4-1}

\end{document}